\documentstyle[12pt]{article}

\begin{document}


\thispagestyle{empty}

\def\pubnum{323}
\def\data{November, 1993}
\def\BB{hep-ph/9402341}

\begin{flushright}
{\parbox{3.5cm}{\rm UAB-FT-\pubnum\\
                \data\\
                \BB}}
\end{flushright}

\vspace{4cm}

\hyphenation{cano-nical ca-nonical canoni-cal}

\begin{center}
\begin{large}
\begin{bf}
ELECTROWEAK SUPERSYMMETRIC QUANTUM CORRECTIONS TO THE TOP QUARK WIDTH\\
\end{bf}
\end{large}
\vspace{1cm}
David GARCIA
\footnote{Internet address:GARCIA@IFAE.ES}\,,
Ricardo A. JIM\'ENEZ
\footnote{Internet address:ABRAHAM@IFAE.ES}\,,
Joan SOL\`A
\footnote{Internet addresses:IFTESOL@CC.UAB.ES and SOLA@IFAE.ES}

\vspace{0.25cm}
Grup de F\'{\i}sica Te\`orica\\

and\\

Institut de F\'\i sica d'Altes Energies\\

\vspace{0.25cm}
Universitat Aut\`onoma de Barcelona\\
08193 Bellaterra (Barcelona), Catalonia, Spain\\
\vspace{1cm}
Wolfgang HOLLIK\\
Institut f\"ur Theoretische Physik,\\
 Universit\"at Karlsruhe,
D-76128 Karlsruhe, Germany\\
%
%
\end{center}
\vspace{0.3cm}
\begin{center}
{\bf ABSTRACT}
\end{center}
\begin{quotation}
\noindent
Within the framework of the MSSM, we compute the electroweak
one-loop supersymmetric quantum corrections to the width
$\Gamma (t\rightarrow W^{+}\, b)$
of the canonical main decay of the top quark.
The results are presented
in two on-shell renormalization schemes parametrized
 either by $\alpha$ or $G_F$.
While in the standard model, and in the Higgs sector of the MSSM, the
electroweak radiative corrections in the $G_F$-scheme
are rather insensitive to the top quark mass and are
 of order of $1\%$ at most, the rest (``genuine'' part) of
the supersymmetric quantum effects in the MSSM
 amount to a non-negligible correction
that could be about one order of magnitude larger, depending on
the top quark mass and
of the region of the supersymmetric parameter space. These new
electroweak effects, therefore,
could be of the same order (and go in the same direction) as
the conventional leading QCD corrections.
\end{quotation}

\baselineskip=6.5mm  


\newpage

\begin{Large}
{\bf 1. Introduction}
\end{Large}
\vspace{0.5cm}

The top quark and the Higgs boson share the priviledge of being the
last two building blocks that remain to be found experimentally
to confirm the fundamental spectrum of the Standard Model (SM)\,
\cite{b001}, and
as such the theoretical consistency of the model
heavily hinges on the existence of these two particles. The replication
of the doublet/singlet pattern structure of the first two
fermion families is required for the suppression of the FCNC in $B$-meson
decays\,\cite{b002}. Moreover, there is a lot of indirect experimental
evidence on the existence of the weak isospin partner of the bottom quark.
The isospin quantum numbers of the $b$-quark can be directly measured
through the partial $Z$-decay width to $b\bar{b}$ pairs and the
forward-backward asymmetry of $b$-quarks at the $Z$-peak yielding,
within small error bars, $T^3(b_L)=-0.5$\,\cite{b003}.
Similarly, although with much lesser accuracy,
the isospin of the RH-component is compatible with zero\,\cite{b003}.

In spite of being a sequential fermion, the top quark plays a special role
in the fermion families due to its huge mass $m_t$
Primordially, the SM predicts a comparatively strong direct interaction
with the Higgs sector through a large Yukawa coupling
\,\cite{b004}.
We thus may expect the top quark as a particularly helpful laboratory
for testing the symmetry breaking mechanism of the SM.
It may also help to unravel
physical effects beyond the SM, such as e.g. those predicted by the Minimal
Supersymmetric Standard Model (MSSM)\,\cite{b005}.
 Direct searches at Tevatron, put a limit of $m_t>113\,GeV$\,\cite{b006},
whereas the combined electroweak data fits from LEP (in the pure
context of the SM) predict
\footnote{From the analysis of $\Delta r$ in the MSSM, it
follows that $m_t$ could be $\sim 20-30\, GeV$ lighter than expected from
the SM, as shown in ref.\cite{b007}}\,\cite{b008}
\begin{equation}
m_t=166\pm 18\pm 21\,GeV \,,
\end{equation}
where the first error is due to measurement errors, while the second arises
 from the uncertainty in the Higgs mass, taken to be between $60\,GeV$ and
$1\, TeV$.
Thus, all phenomenological evidence points towards the top quark being
around the corner, and may be within the discovery potential of the
Tevatron ($m_t\leq 180\,GeV$). At the hadron-hadron supercollider LHC and a
next linear $e^{+}e^{-}$ collider, the $t\bar{t}$ system
will be copiously produced through parton fusion and $e^{+}e^{-}$
anihilation, respectively, and the decay modes analyzed in great detail\,
\cite{b009,b010}.
Therefore, precise measurements of the top quark properties
will become available facing the predictions of the SM, and we should be
prepared to recognize or to
exclude hints of new physics. Notice that for $m_t\geq 130\, GeV$, the
width $\Gamma_t\equiv \Gamma (t\rightarrow W^{+}\, b)$ exceeds
$\sim 0.5\, GeV$ and thus
$\Gamma_t>\Lambda_{QCD}$. As a consequence, the top quark will predominantly
decay as a free quark, and the bound states cannot be formed,
leading to a broad threshold enhancement in the production process
of $t\bar{t}$ pairs instead of sharp resonances\,\cite{b310}.
 This allows to analyze
the production and decay of top quarks perturbatively, with $\Gamma_t$
serving as the infrared cutoff \,\cite{b210}.

Radiative corrections to conventional
physical processes\,\cite{b110}
are a powerful tool to search for mass scales within and
beyond the SM, and they offer us the opportunity to peep at sectors
of the theory that are not (yet) directly observable.
In this paper we concentrate on the computation of the
supersymmetric (SUSY) quantum effects on the width of the canonical decay
$t\rightarrow W^{+}\, b$, probably
the main decay mode of the top quark. In extended versions of the SM
(Cf.sect.4) other decay channels may also be open, but the standard decay
always has a sizeable branching ratio.
In the framework of the SM, the aforementioned limit
on the top quark mass is based on that standard decay and
the tagging of the subsequent leptonic decay mode of the weak gauge
boson, with an approximate branching ratio of
$BR(t\rightarrow l\,\nu_l\,b)\simeq 1/9$.
Detailed analyses of the electroweak one-loop effects
on the canonical decay, in the pure
context of the SM, already exist in the literature,
with the result (somewhat surprising) that they are of order of $1-1.5\%$
at most and they turn out to be rather insensitive to the top quark mass in
the relevant range $100-200\,GeV$\,\cite{b211,b013,b213}.

The motivation of this calculation are the potentially large quantum effects
on the top decay width arising from extra significant interactions between
heavy fermions and the Higgs sector.
In supersymmetric
extensions of the SM, the Higgs sector contains
at least two superfield Higgs doublets.
The corresponding analysis for two-doublet Higgs
(SUSY and non-SUSY) extensions of the SM was first given in
ref.\cite{b014}, with
the result that no large corrections ($\le 1\%$)
on the top quark width are gained from the SUSY scalar
Higgses alone. Notwithstanding, a full account of the remaining
--``genuine''-- SUSY contribution: namely, from sfermions
 (squarks and sleptons) and
``inos'' (charginos and neutralinos), was still missing, but the relevant
Higgs-like interactions involving these two set of  supersymmetric particles
(``sparticles'') provide another source of large loop contributions,
in particular if the sparticles are not too heavy.
Such an interesting situation of a ``light'' effective low energy
SUSY scale $M_{SUSY}$ (i.e. that scale fixed by the renormalized
soft SUSY-breaking terms) around the Fermi scale (or even below)
is compatible with the intriguing coupling constant unification
in a SUSY-GUT scenario\,\cite{b015} consistent with the
high precision LEP data and the non-observation of proton decay\,
\cite{b016,b017}. In this paper we exploit the possibility to obtain
indirect information both on SUSY physics and on top quark dynamics.

The paper is organized as follows. In sect.2 we present a quick review of
the basic SUSY formalism necessary for our calculation and give those
parts of the interaction Lagrangian describing the
fermion-sfermion-chargino/neutralino coupling.  In sect.3 we display
the results of the analytical calculation of the various
electroweak one-loop MSSM
contributions to the top quark decay width within the framework of the
standard on-shell renormalization framework. Finally, sect.4 is
devoted to a detailed presentation of
the numerical analysis and a corresponding discussion of the results.


\vspace{1cm}
\hyphenation{diago-nalization diagona-lization}
\begin{Large}
{\bf 2. SUSY Formalism and Interaction Lagrangian}
\end{Large}
\vspace{0.5cm}

We shall perform our calculations in a mass-eigenstate basis.
One goes from
the weak-eigenstate basis to the mass-eigenstate basis via appropriate
unitary transformations. Two classes of SUSY particles enter our
calculations: the fermionic partners of gauge bosons and Higgs bosons
(called gauginos, $\tilde{W}$, and higgsinos, $\tilde{H}$, respectively)
and on the other hand the scalar partners of quarks
and leptons
(called squarks, $\tilde{q}$,  and sleptons, $\tilde{l}$,
respectively, or  sfermions, $\tilde{f}$,
generically). Within the context of the MSSM,
we need two Higgs superfield doublets with weak hypercharges $Y_{1,2}=\mp 1$
( hats denote superfields):
\begin{equation}
 \hat{H}_1= \left(\begin{array}{c}
\hat{H}_1^0 \\ \hat{H}_1^{-}
\end{array} \right)
 \;\; \;\;\;,\;\;\;\;\;
 \hat{H}_2= \left(\begin{array}{c}
\hat{H}_2^{+} \\ \hat{H}_2^0
\end{array} \right)\ .
\end{equation}
The corresponding scalar Higgs doublet $H_1$ ($H_2$)
gives mass to the down (up) -like quarks
through the VEV $<H^0_1>=v_1$ ($<H^0_2>=v_2$).
This is seen from the structure of the MSSM superpotential\,\cite{b005}
\begin{equation}
\hat{W}=\epsilon_{ij}[h_b\,\hat{H}_1^i\hat{Q}^j\hat{D}+
h_t\hat{H}_2^j\hat{Q}^i\hat{U}+\mu\hat{H}_1^j\hat{H}_2^i]\,,
\label{eq:W}
\end{equation}
where we have singled out only the Yukawa couplings of
the third quark generation, $(t,b)$, as a  generical
fermion-sfermion generation of chiral matter superfields
$\hat{Q}$, $\hat{U}$ and $\hat{D}$.
Their respective scalar (squark) components are:
\begin{equation}
\tilde{Q}=\left(\begin{array}{c}
\tilde{t'}_L \\ \tilde{b'}_L
\end{array} \right) \; \;\; ,\;\;\;\tilde{U}=\tilde{t'}_R^{*}
\; \;\; ,\;\;\;\tilde{D}=\tilde{b'}_R^{*}\,,
\label{eq:QUD}
\end{equation}
with weak hypercharges $Y_Q=+1/3$, $Y_U=-4/3$ and $Y_D=+2/3$.
The primes in (\ref{eq:QUD}) denote the fact that
$\tilde{q'}_a=\{\tilde{q'}_L, \tilde{q'}_R\}$ are weak-eigenstates, not
mass-eigenstates. However, there
may be  ``chiral'' L-R mixing between weak-eigenstate
sfermions of a given flavor (except
for the sneutrinos), which is induced already
at tree-level by the $\mu$-term in the
superpotential and by the (renormalized) trilinear ``soft''
SUSY-breaking terms \,\cite{b005}.
Due to this mixing, we have to derive the corresponding
squark mass-eigenstates $\tilde{q}_a=\{\tilde{q}_1, \tilde{q}_2\}$
by means of appropriate
$2\times 2$ rotation matrices, $ R^{(q)}$,
that diagonalize the chiral mass matrices
(we neglect intergenerational mixing):
\begin{equation}
\tilde{q}_a=\sum_{b} R_{ab}^{(q)}\tilde{q'}_b\;\;\;\;
(\tilde{q}=\tilde{t}, \tilde{b}) .
\end{equation}
 From the higgsinos and the various gauginos we form
the following three sets of two-component Weyl spinors:
 \begin{equation}
\Gamma_i^{+} = \{-i\tilde{W}^{+}, \tilde{H_2^{+}}\}\ , \;\;\;
\Gamma_i^{-} = \{-i\tilde{W}^{-}, \tilde{H_1^{-}}\}\ ,
\end{equation}
\begin{equation}
\Gamma_{\alpha}^{0} =
\{-i\tilde{B^{0}}, -i\tilde{W_3^{0}}, \tilde{H_2^0}, \tilde{H_1^0}\}\ .
\end{equation}
These states get mixed up when the neutral Higgs fields acquire nonvanishing
VEV's giving masses to the gauge bosons:
$M_W^2=(1/2) g^2 (v_1^2+v_2^2),\, M_Z^2=(1/2) (g^2+g'^2) (v_1^2+v_2^2)$.
The ``ino'' mass Lagrangian reads
\begin{equation}
{\cal L}_M=-<\Gamma^{+}|{\cal M}|\Gamma^{-}>-{1\over2}
<\Gamma^0|{\cal M}^0|\Gamma^0>+h.c.\ ,
\end{equation}
where the charged and neutral gaugino-higgsino mass
matrices are the following:
\begin{equation}
{\cal M} =\left(\begin{array}{cc}
M  &  M_W\sqrt{2}\,c_{\beta} \\
M_W\sqrt{2}\,s_{\beta} & \mu
\end{array} \right)
\end{equation}
and
\begin{equation}
{\cal M}^0 =\left(\begin{array}{cccc}
M' & 0 & M_Z\,s_{\beta}\,s_{\theta} & -M_Z\,c_{\beta}\,s_{\theta} \\
0 & M & -M_Z\,s_{\beta}\,c_{\theta} & -M_Z\,c_{\beta}\,c_{\theta} \\
M_Z\,s_{\beta}\,s_{\theta} & -M_Z\,s_{\beta}\,c_{\theta}&0&-\mu \\
-M_Z\,c_{\beta}\,s_{\theta} &M_Z\,c_{\beta}\,c_{\theta}&-\mu&0
\end{array} \right)\ ,
\end{equation}
with the following notation:
\begin{equation}
s_{\beta}\equiv\sin{\beta}\;\;\ ,\;\; c_{\beta}\equiv\cos{\beta}\;\;\ ,\;\;
\tan{\beta}={v_2\over v_1}\;\;\ ,\;\;c_{\theta}\equiv M_W/M_Z
\;\;\ ,\;\;s_{\theta}^2\equiv 1-c_{\theta}^2 .
\label{eq:angles}
\end{equation}
The mass parameters $M$ and $M'$ come from $SU(2)_L\times U(1)_Y$-invariant
gaugino mass terms that softly break
global SUSY,
while $\mu$ is the very same SUSY Higgs mass term in the superpotential
(\ref{eq:W}).
We shall assume that the MSSM can be embedded in a GUT,
in which case the parameters $M'$ and $M$ are related as follows
\,\cite{b005,b015}
\begin{equation}
{M'\over M}={5\over 3}\,t_{\theta}^2\simeq 0.5\,,
\end{equation}
where  $t_{\theta}\equiv s_{\theta}/c_{\theta}$.
The $2\times 2$ mass matrix ${\cal M}$ is in general non-symmetrical and its
diagonalization is accomplished by two unitary matrices $U$ and $V$, whereas
the symmetrical $4\times 4$ mass matrix ${\cal M}^0$ can be diagonalized by
a single unitary matrix $N$:
\begin{equation}
U^{*}{\cal M}V^{\dagger}=diag\{M_1,M_2\}\;\;\;\;\;,\;\;\;\;\;
N^{*}{\cal M}^0N^{\dagger}=diag\{M_1^0,...M_4^0\}\ .
\end{equation}
 Let us now build the charged mass-eigenstate 4-spinors (charginos)
associated to the mass eigenvalues $M_i$. Call them $\Psi_i^+$, and
let $\Psi_i^-$ be the corresponding charge conjugate states.
We have\footnote{We use the notation of ref.\cite{b018} for the sparticles
and their indices. We remark that
first Latin indices a,b,...=1,2 are reserved for sfermions,
middle Latin indices i,j,...=1,2
for charginos, and first Greek
indices $\alpha, \beta,...=1,2,...,4$ for neutralinos. }
\begin{equation}
 \Psi_i^{+}= \left(\begin{array}{c}
U_{ij}\Gamma_j^{+} \\ V_{ij}^{*}\bar{\Gamma}_j^{-}
\end{array} \right)
 \; \;\; \;\;,\;\;\;\;\;
 \Psi_i^{-}= C\bar{\Psi_i}^{-T} =\left(\begin{array}{c}
V_{ij}\Gamma^{-}_j \\ U_{ij}^{*}\bar{\Gamma}_j^{+}
\end{array} \right)\ .
\end{equation}
As for the neutral mass-eigenstate 4-spinors (neutralinos) associated to the
mass eigenvalues $M_{\alpha}^0$, they are the following
Majorana spinors
\begin{equation}
 \Psi_{\alpha}^0= \left(\begin{array}{c}
N_{\alpha,\beta}\Gamma_{\beta}^0 \\
N_{\alpha,\beta}^{*}\bar{\Gamma}_{\beta}^0
\end{array} \right) =
C\bar{\Psi}_{\alpha}^{0T}\ .
\end{equation}
The process-dependent SUSY diagrams contributing to\,
$t\rightarrow W^{+}\,b$\, include only a limited portion
of the MSSM Lagrangian.  On the other hand, the computation
of the (universal) counterterms (Cf. sect.3) associated to the
on-shell renormalization
procedure does require the use of the full electroweak SUSY part.
We refer the reader to the literature for the remaining structure
of the MSSM Lagrangian\,\cite{b005,b015,b018}.
Here, however, we shall exhibite
explicitly only that part of the interaction Lagrangian needed for the
computation of the specific one-loop vertices related to our process,
emphasizing that part with the relevant
Yukawa couplings in the mass-eigenstate basis.
In order to construct it, we project the quark-squark-higgsino terms from the
superpotential (\ref{eq:W}). Furthermore,
there are also gaugino interactions that mix with
these terms; they come from expanding the SUSY counterpart of the
 $SU(2)_L\times U(1)_Y$ fermion-gauge interaction terms, viz.
\begin{equation}
{\cal L}_{\tilde{q}\lambda\,q}=i\sqrt{2}\,g_r\,
\tilde{q}^{*}T^r\,\lambda^r\,q+h.c.\,,
\end{equation}
with
\begin{equation}
\lambda^r=\{\stackrel{\rightarrow}{\tilde{W}}, \tilde{B}\}\;\;\;,\;\;\;
T^r=\{\stackrel{\rightarrow}{\sigma}/2,Y/2\}\;\;\;,
\;\;\;g_r=g,g'\,.
\end{equation}
After addying up these two kinds of terms in the weak-eigenstate,
two-component, basis and re-expressing the result in
four-component notation and in the mass-eigenstate basis we find
\begin{eqnarray}
{\cal L}_{\Psi\,q\,\tilde{q}} &=&\sum_{a=1,2}\sum_{i=1,2}\left\{
-g\,\tilde{t}_a^{*}\,\bar{\Psi}_i^{-}\,{1\over 2}\,
(A_{ai}^{(t)}-B_{ai}^{(t)}\,\gamma_5)\,b
-g\,\tilde{b}_a^{*}\,\bar{\Psi}_i^{+}\,{1\over 2}\,
(A_{ai}^{(b)}-B_{ai}^{(b)}\,\gamma_5)\,t\right\}\nonumber\\
& &\sum_{a=1,2}\sum_{\alpha=1,...,4}\left\{
-{g\over \sqrt{2}}\,\tilde{t}_a^{*}\,\bar{\Psi}_{\alpha}^{0}\,{1\over 2}\,
(A_{a\alpha}^{(t)}-B_{a\alpha}^{(t)}\,\gamma_5)\,t
+{g\over \sqrt{2}}\,\tilde{b}_a^{*}\,\bar{\Psi}_{\alpha}^{0}\,{1\over 2}\,
(A_{a\alpha}^{(b)}-B_{a\alpha}^{(b)}\,\gamma_5)\,b\right\}\nonumber\\
& & +h.c.\,,
\end{eqnarray}
where, using the notation introduced above,
we have defined the
following coupling matrices:
\begin{eqnarray}
A_{ai}^{(t)}&=& R_{a1}^{(t)}\,(U_{i1}^{*}-
\lambda_b\,V_{i2})-\lambda_t\,R_{a2}^{(t)}\,U_{i2}^{*}\,,\nonumber\\
B_{ai}^{(t)}&=& R_{a1}^{(t)}\,(U_{i1}^{*}+
\lambda_b\,V_{i2})-\lambda_t\,R_{a2}^{(t)}\,U_{i2}^{*}\,,
\nonumber\\
\nonumber\\
A_{ai}^{(b)}&=& R_{a1}^{(b)}\,(V_{i1}^{*}-
\lambda_t\,U_{i2})-\lambda_b\,R_{a2}^{(b)}\,V_{i2}^{*}\,,\nonumber\\
B_{ai}^{(b)}&=& R_{a1}^{(b)}\,(V_{i1}^{*}+
\lambda_t\,U_{i2})-\lambda_b\,R_{a2}^{(b)}\,V_{i2}^{*}\,,
\nonumber\\
\nonumber\\
A_{a\alpha}^{(t)}&=& R_{a1}^{(t)}\,(N_{\alpha 2}^{*}+
{1\over 3}\,t_{\theta}\,N_{\alpha 1}^{*}+\sqrt{2}\,\lambda_t\,N_{\alpha 3})
-R_{a2}^{(t)}\,({4\over 3}\,t_{\theta}\,N_{\alpha 1}
-\sqrt{2}\,\lambda_t\,N_{\alpha 3}^{*})\,,\nonumber\\
B_{a\alpha}^{(t)}&=& R_{a1}^{(t)}\,(N_{\alpha 2}^{*}+
{1\over 3}\,t_{\theta}\,N_{\alpha 1}^{*}-\sqrt{2}\,\lambda_t\,N_{\alpha 3})
+R_{a2}^{(t)}\,({4\over 3}\,t_{\theta}\,N_{\alpha 1}
+\sqrt{2}\,\lambda_t\,N_{\alpha 3}^{*})\,,
\nonumber\\
\nonumber\\
A_{a\alpha}^{(b)}&=& R_{a1}^{(b)}\,(N_{\alpha 2}^{*}-
{1\over 3}\,t_{\theta}\,N_{\alpha 1}^{*}-\sqrt{2}\,\lambda_b\,N_{\alpha 4})
-R_{a2}^{(b)}\,({2\over 3}\,t_{\theta}\,N_{\alpha 1}
+\sqrt{2}\,\lambda_b\,N_{\alpha 4}^{*})\,,\nonumber\\
B_{a\alpha}^{(b)}&=& R_{a1}^{(b)}\,(N_{\alpha 2}^{*}-
{1\over 3}\,t_{\theta}\,N_{\alpha 1}^{*}+\sqrt{2}\,\lambda_b\,N_{\alpha 4})
+R_{a2}^{(b)}\,({2\over 3}\,t_{\theta}\,N_{\alpha 1}
-\sqrt{2}\,\lambda_b\,N_{\alpha 4}^{*})\,.
\end{eqnarray}
The potentially significant Yukawa couplings are contained in the
following ratios with respect to the $SU(2)_L$ gauge coupling:
\begin{equation}
\lambda_t\equiv{h_t\over g}={m_t\over \sqrt{2}\,M_W\,s_{\beta}}\;\;\;\;\;,
\;\;\;\;\;\lambda_b\equiv{h_b\over g}={m_b\over \sqrt{2}\,M_W\,c_{\beta}}\,.
\label{eq:Yukawas}
\end{equation}
Finally, the relevant charged-current interaction of squarks
and charginos with the $W^{\pm}$
gauge bosons is given by
\begin{eqnarray}
{\cal L}_W^{CC} & = &{ig\over \sqrt{2}}
\sum_{a,c}\{R_{a1}^{(b)*} R_{c1}^{(t)}\,\tilde{t}_c^{*}
\stackrel{\leftrightarrow}{\partial^{\mu}}\tilde{b}_aW_{\mu}^{+}\}
 +  J_{+}^{\mu}W_{\mu}^{-}
+h.c.\nonumber\\
& + & {1\over 2}g^2\sum_{a,c}\{R_{a1}^{(t)*} R_{c1}^{(t)}
\,\tilde{t}_c^{*}\tilde{t}_a+
R_{a1}^{(b)*} R_{c1}^{(b)}
\,\tilde{b}_c^{*}\tilde{b}_a\}W_{\mu}^{+}W^{-\mu}\,,
\end{eqnarray}
where
\begin{equation}
 J_{+}^{\mu}=g\sum_{\alpha}\sum_i\bar{\Psi}^0_{\alpha}\gamma^{\mu}
(C_{\alpha i}^LP_L+C_{\alpha i}^RP_R){\Psi}_i^{+}\ ,
\end{equation}
with $P_{L,R}=(1/2)(1\pm\gamma_5)$, and the chargino coupling matrices
\begin{eqnarray}
C_{\alpha i}^L &=& {1\over \sqrt{2}}N_{\alpha 3}U_{i2}^{*}
-N_{\alpha 2}U_{i1}^{*}\nonumber\\
C_{\alpha i}^R &=& {-1\over \sqrt{2}}N_{\alpha 4}^{*}V_{i2}
-N_{\alpha 2}^{*}V_{i1}\ .
\end{eqnarray}


\vspace{1.5cm}
\begin{Large}
\begin{quote}
\raggedright
{\bf 3. Supersymmetric Quantum Corrections  }
\end{quote}
\end{Large}
\hyphenation{nume-rical numeri-cal}
 \vspace{0.5cm}

In our calculation of the one-loop electroweak corrections to
 $\Gamma_t\equiv\Gamma (t\rightarrow W^{+}\, b)$ in the MSSM, we shall adopt
the on-shell renormalization scheme\,\cite{b019},
where the fine structure constant,
$\alpha$, and the masses of the gauge bosons, fermions and scalars are
the renormalized parameters: $(\alpha, M_W, M_Z, M_H, m_f, M_{SUSY},...)$
\footnote{ For a comprehensive review, see e.g. refs.
\cite{b110,b020,b021}.}. We will, for brevity sake,
refer to it as the ``$\alpha$-scheme'': $(\alpha, M_W, M_Z)$.
As stated, the corrections to\, $\Gamma_t$\, from a general
two-Higgs-doublet model (2HDM), and in particular that of the MSSM,
were already considered in ref.\cite{b014} within the framework of the
``minimal'' $\alpha$-scheme of ref.\cite{b020}
\footnote{See also the alternative, though fully equivalent, calculation
of ref.\cite{b022} within the framework of the  ``complete''
(matrix) $\alpha$-scheme of ref.\cite{b021}.},
which we shall
also adhere to in this work.
We shall therefore concentrate on the
remaining supersymmetric electroweak corrections: namely, from charginos,
neutralinos and sfermions. The direct vertex corrections originating from
this ``genuine'' supersymmetric sector of the MSSM are depicted in Fig.1.
The bare structure of any of these vertices can be separated as the
sum of the tree-level part plus one-loop correction:
\begin{equation}
\Gamma_{\mu}^{(0)}=i\,{g\over \sqrt{2}}[\gamma_{\mu}\,P_L(1+F_L)+
\gamma_{\mu}\,P_R\,F_R+{p_{\mu}\over M_W}(P_L\,H_L+P_R\,H_R)]\,,
\label{eq:G1}
\end{equation}
where the correction has been parametrized in terms of four form factors
$F_L, F_R, H_L$ and
$ H_R$,
of which only $F_L$ is UV-divergent\,\cite{b014}.
The corresponding renormalized vertex
$\Gamma^{\mu}\rightarrow \Gamma^{\mu}+\delta \Gamma^{\mu}$
is obtained from the renormalized Lagrangian
${\cal L}\rightarrow {\cal L}+\delta{\cal L}$
in the on-shell renormalization framework. In the minimal
$\alpha$-scheme of ref.\cite{b020},
where a minimum number of field renormalization
constants is used (viz. one renormalization constant per symmetry multiplet),
the effect of the counterterm Lagrangian $\delta{\cal L}$ is equivalent
to the following replacement of the UV-divergent form factor
in eq.(\ref{eq:G1}):
\begin{equation}
F_L\rightarrow F_L+ \delta Z_1^W-\delta Z_2^W +
\delta Z_L\,,
\end{equation}
and the resulting expression has to be finite.
 Here $Z_i=1+\delta Z_i$
are the renormalization constants
defined by\,\cite{b020}
\begin{eqnarray}
W_{\mu}&\rightarrow & {(Z_2^W)}^{1/2}\,W_{\mu}\nonumber\\
\left(\begin{array}{c}
t_L \\ b_L
\end{array} \right)
&\rightarrow & Z_L^{1/2}\,
\left(\begin{array}{c}
t_L \\ b_L
\end{array} \right)
\nonumber\\
g &\rightarrow & {(Z_1^W)}\,{(Z_2^W)}^{-3/2}\, g\,.
\end{eqnarray}
Explicitly they read
as follows
\begin{eqnarray}
\delta Z_1^W & = &
\left.  {\Sigma^{\gamma}(k^2)\over k^2}\right|_{k^2=0}+{1+2\,c_{\theta}^2
\over s_{\theta}\,c_{\theta}}\,{\Sigma^{\gamma Z}(0)\over M_Z^2}
+{c_{\theta}^2\over s_{\theta}^2}
({\delta M_Z^2\over M_Z^2}-{\delta M_W^2\over M_W^2})
\nonumber\\
\delta Z_2^W & = &
\left.  {\Sigma^{\gamma}(k^2)\over k^2}\right|_{k^2=0}+2{c_{\theta}
\over s_{\theta}}\,{\Sigma^{\gamma Z}(0)\over M_Z^2}
+{c_{\theta}^2\over s_{\theta}^2}
({\delta M_Z^2\over M_Z^2}-{\delta M_W^2\over M_W^2})\nonumber\\
\delta Z_L &=& \Sigma^b_L(m_b^2)+m_b^2[\Sigma^{b\,\prime}_L
(m_b^2)+\Sigma^{b\,\prime}_R(m_b^2)
+2\Sigma^{b\,\prime}_S(m_b^2)]\,,
\label{eq:deltaZ}
\end{eqnarray}
where $\Sigma^{\gamma, W, Z,...}(k^2)$
are the real parts of the
various (transverse components of the) gauge boson self-energy
functions \footnote{Our self-energy functions\,\cite{b018}
are opposite in sign to those of ref.\cite{b020}.}.
The gauge boson mass counterterms
\begin{equation}
\delta M_W^2=-Re\,\Sigma^W(k^2=M_W^2)\,\,,\ \ \ \ \
\delta M_Z^2=-Re\,\Sigma^Z(k^2=M_Z^2)
\label{eq:MCT}
\end{equation}
are enforced by the on-shell renormalization conditions. Moreover, we have
decomposed the (real part of the) bottom quark self-energy according to
\begin{equation}
\Sigma^b(p)=\Sigma^b_L(p^2)\not{p}\,P_L+\Sigma^b_R(p^2)\not{p}\,P_R
+m_b\,\Sigma^b_S(p^2)\,,
\label{eq:Sigma}
\end{equation}
and used the notation $\Sigma'(p)\equiv \partial\Sigma(p)/\partial p^2$.
Notice that in the minimal $\alpha$-scheme of ref.\cite{b020}, where a single
renormalization constant is assigned to the quark doublet
$(t,b)$, it is impossible
to arrange for the residues of the top and bottom quark propagators to be
simultaneously equal to one. In our case only the bottom quark propagator is
normalized this way.
Consequently, one is forced to introduce a finite
wave-function renormalization for the top quark external line (Cf. Fig.2):
${1\over 2}\hat{\Pi}_t(m_t^2)$, where
\begin{eqnarray}
\hat{\Pi}_t(m_t^2) &=&{\Pi_t}(m_t^2)+\delta Z_L\nonumber\\
{\Pi_t}(m_t^2) &=& \Sigma^t_L(m_t^2)+m_t^2
[\Sigma^{t\,\prime}_L(m_t^2)\not{p}P_L+
\Sigma^{t\,\prime}_R(m_t^2)\not{p}P_R
+2m_t\Sigma^{t\,\prime}_S(m_t^2)]\,.
\label{eq:pito}
\end{eqnarray}
Similarly,  ${1\over 2}\hat{\Sigma}^{'}_W(M_W^2)$ gives the finite
wave-function renormalization of the external $W$ (Cf. Fig.3),
where the renormalized $W$-self-energy is given by
\begin{equation}
\hat{\Sigma}_W(k^2)={\Sigma}_W(k^2)-\delta M_W^2+\delta Z_2^W(k^2-M_W^2)
\label{eq:RSEW}
\end{equation}
and so
\begin{equation}
\hat{\Sigma}_W'(M_W^2)={\Sigma}_W'(M_W^2)+\delta Z_2^W\,.
\label{eq:DRSEW}
\end{equation}
Putting things together, the general structure of the (minimally) on-shell
renormalized $t\,b\,W$-vertex is
\begin{eqnarray}
\Gamma_{\mu} &=& \Gamma_{\mu}^{(0)}+\delta\Gamma_{\mu}\nonumber\\
\delta\Gamma_{\mu} &=& i\,{g\over \sqrt{2}}\gamma_{\mu}\,P_L[\delta Z_L
+\delta Z_1^W-\delta Z_2^W
+{1\over 2}\hat{\Pi}_t(m_t^2)+{1\over 2}\hat{\Sigma}_W'(M_W^2)]\,.
\label{eq:RV}
\end{eqnarray}
In this expression, the full SUSY pay-off
to the combined counterterm $\delta Z_1^W-\delta Z_2^W $ turns out to vanish,
for the latter combination is seen from eq.(\ref{eq:deltaZ}) to be
proportional to the mixed self-energy function
$\Sigma^{\gamma Z}(k^2)$ at $k^2=0$,
and all the chargino-neutralino and sfermion contributions to this function
identically vanish at zero frequency\,\cite{b018}.

We are now ready to compute the $48$ one-loop SUSY contributions from the
diagrams of Fig.1 to the vertex form factors. The computation is rather
involved, since we retain exact dependence on all masses and keep track of
all matrix coupling constants for all SUSY particles in their respective
mass-eigenstate bases. Nevertheless, we have managed to present the
final analytical results in a fairly compact form.
Apart from the tree-level diagram v0 in Fig.1,
there are three basic one-loop vertex
diagrams: v1, v2 and v3, and all of them are summed over all ``ino''
and sfermion indices according to the notation of sect.2, which we shall use
extensively hereafter.

{\bf Diagram v1}:  Define the following matrices
\footnote{Matrix indices are always positioned as lower indices,
whereas parenthetical upper indices are reserved to denote either
flavor (f)=(b),(t) or, in case of (0), to refer to neutralino
coupling matrices, when necessary.}
\begin{equation}
A_{\pm }\equiv A_{\pm ai}^{(t)}
=A_{ai}^{(t)}\pm B_{ai}^{(t)}\,,\ \ \ \ \
A_{\pm}^{(0)}\equiv A_{\pm a\alpha}^{(t)}
=A_{a\alpha}^{(t)}\pm B_{a\alpha}^{(t)}
\label{eq:A11}
\end{equation}
and construct (omitting all indices for simplicity) the combinations
\begin{eqnarray}
{\bf A}^{(1)}&=& A_{+}^{*}C^R A_{-}^{(0)}\, \ \ \ \ \ \ \ \
{\bf E}^{(1)} = A_{-}^{*}C^R A_{-}^{(0)}\nonumber\\
{\bf B}^{(1)}&=& A_{+}^{*}C^R A_{+}^{(0)}\, \ \ \ \ \ \ \ \
{\bf F}^{(1)} = A_{-}^{*}C^R A_{+}^{(0)}\nonumber\\
{\bf C}^{(1)}&=& A_{+}^{*}C^L A_{-}^{(0)}\, \ \ \ \ \ \ \ \
{\bf G}^{(1)} = A_{-}^{*}C^L A_{-}^{(0)}\nonumber\\
{\bf D}^{(1)}&=& A_{+}^{*}C^L A_{+}^{(0)}\, \ \ \ \ \ \ \ \
{\bf H}^{(1)} = A_{-}^{*}C^L A_{+}^{(0)}\,.
\label{eq:A12}
\end{eqnarray}
Then the contribution from
diagram v1 to the form factors in eq.(\ref{eq:G1}) is the following:
\begin{eqnarray}
F_L^{(v1)} &=& -{ig^2\over 4}[-{\bf D}^{(1)}\tilde{C}_0+M_W^2{\bf D}^{(1)}
(C_0+C_{11}-C_{12})+m_t^2{\bf D}^{(1)}(-C_0-2C_{11}+2C_{12})\nonumber\\
& + & m_b^2{\bf D}^{(1)}(-C_0-C_{11})
 +  m_t\,m_b{\bf E}^{(1)}(-C_0-C_{11}+2C_{12})\nonumber\\
& + & m_t\,M_i{\bf A}^{(1)}
(C_0+C_{11}-C_{12})
 +  m_t\,M_{\alpha}^0{\bf C}^{(1)}(-C_{11}+C_{12})
 -  m_b\,M_i{\bf H}^{(1)}C_{12}\nonumber\\
& + & m_b\,M_{\alpha}^0{\bf F}^{(1)}
(C_0+C_{12})
 +  M_i\,M_{\alpha}^0{\bf B}^{(1)}C_0 +2{\bf D}^{(1)}C_{24}]\,,
\nonumber\\
F_R^{(v1)} &=& -{ig^2\over 4}[-{\bf E}^{(1)}\tilde{C}_0+M_W^2{\bf E}^{(1)}
(C_0+C_{11})+m_t^2{\bf E}^{(1)}(-C_0-2C_{11}+C_{12})\nonumber\\
& + & m_b^2{\bf E}^{(1)}(-C_0-C_{11}-C_{12})
 +  m_t\,m_b{\bf D}^{(1)}(-C_0-C_{11})\nonumber\\
& + & m_t\,M_i{\bf H}^{(1)}
(C_0+C_{11}-C_{12})
 +  m_t\,M_{\alpha}^0{\bf F}^{(1)}(-C_{11}+C_{12})
 -  m_b\,M_i{\bf A}^{(1)}C_{12}\nonumber\\
& + & m_b\,M_{\alpha}^0{\bf C}^{(1)}(C_0+C_{12})
 +  M_i\,M_{\alpha}^0{\bf G}^{(1)}C_0 +2{\bf E}^{(1)}C_{24}]\,,
\nonumber\\
H_L^{(v1)} &=& -{ig^2 M_W\over 2}[m_t{\bf E}^{(1)}
(2C_0+3C_{11}-3C_{12}+C_{21}-C_{23})
+m_b{\bf D}^{(1)}(C_{12}+C_{23})\nonumber\\
& + & M_i{\bf H}^{(1)}C_{12}
 +  M_{\alpha}^0{\bf F}^{(1)}(C_{11}-C_{12})]\,,
\nonumber\\
H_R^{(v1)} &=& -{ig^2 M_W\over 2}[m_t{\bf D}^{(1)}
(2C_0+3C_{11}-3C_{12}+C_{21}-C_{23})
+m_b{\bf E}^{(1)}(C_{12}+C_{23})\nonumber\\
& + & M_i{\bf A}^{(1)}C_{12}
+M_{\alpha}^0{\bf C}^{(1)}(C_{11}-C_{12})]\,.
\label{eq:V1}
\end{eqnarray}
In the previous expressions we have used (and also explicitly checked)
the 3-point function notation from
ref.\cite{b023},
which is an adaptation to the $g_{\mu\nu}=(-,+,+,+)$ metric
of the standard formulae from refs.\cite{b024,b025}.
For diagram v1, all the 3-point functions
($C$'s and $\tilde{C}_0$) in (\ref{eq:V1}) have the arguments
\begin{equation}
C=C(p^2,p'^2,m_a(\tilde{t}), M_{\alpha}^0, M_i)\,.
\end{equation}

{\bf Diagram v2}: Define the following matrices:
\begin{equation}
A_{\pm}^{(b)}\equiv A_{\pm a\alpha}^{(b)}
=A_{a\alpha}^{(b)}\pm B_{a\alpha}^{(b)}\,,\ \ \ \ \
A_{\pm}^{(t)}\equiv A_{\pm a\alpha}^{(t)}
=A_{a\alpha}^{(t)}\pm B_{a\alpha}^{(t)}
\label{eq:A21}
\end{equation}
and form the combinations
\begin{eqnarray}
{\bf A}^{(2)}&=& R_{b1}^{(t)*}R_{a1}^{(b)}A_{+}^{(b)*}
A_{-}^{(t)}\, \ \ \ \ \ \ \ \
{\bf C}^{(2)} =  R_{b1}^{(t)*}R_{a1}^{(b)}A_{-}^{(b)*}
A_{-}^{(t)}\nonumber\\
{\bf B}^{(2)}&=& R_{b1}^{(t)*}R_{a1}^{(b)}A_{+}^{(b)*}
A_{+}^{(t)}\, \ \ \ \ \ \ \ \
{\bf D}^{(2)} = R_{b1}^{(t)*}R_{a1}^{(b)}A_{-}^{(b)*}A_{+}^{(t)}\,.
\label{eq:A22}
\end{eqnarray}
The contributions from v2 to the form factors are:
\begin{eqnarray}
F_L^{(v2)} &=& {ig^2\over 4}{\bf B}^{(2)}C_{24}\,,
\nonumber\\
F_R^{(v2)} &=& {ig^2\over 4}{\bf C}^{(2)}C_{24}\,,
\nonumber\\
H_L^{(v2)} &=& {-ig^2 M_W\over 4}[m_t{\bf C}^{(2)}(-C_{11}+C_{12}-C_{21}+
C_{23})+m_b{\bf B}^{(2)}(-C_{12}-C_{23})\nonumber\\
& + & M_{\alpha}^0{\bf D}^{(2)}(C_0+C_{11})\,,
\nonumber\\
H_R^{(v2)} &=& {-ig^2 M_W\over 4}[m_t{\bf B}^{(2)}(-C_{11}+C_{12}-C_{21}+
C_{23})+m_b{\bf C}^{(2)}(-C_{12}-C_{23})\nonumber\\
& + & M_{\alpha}^0{\bf A}^{(2)}(C_0+C_{11})\,.
\label{eq:V2}
\end{eqnarray}
In this case the
3-point functions in eq.(\ref{eq:V2}) have the arguments
\begin{equation}
C=C(p^2,p'^2, M_{\alpha}^0, m_b(\tilde{t}), m_a(\tilde{b}) )\,.
\end{equation}

{\bf Diagram v3}:  The structure of the various contributions is very
similar to those from diagram v1. They can be obtained by just
replacing $M_i\leftrightarrow M_{\alpha}^0$ everywhere in eq.(\ref{eq:V1})
and at the same time substituting the set of matrices
\begin{equation}
A_{\pm}\equiv A_{\pm ai}^{(b)}=A_{ai}^{(b)}\pm B_{ai}^{(b)}\,,\ \ \ \ \
A_{\pm}^{(0)}\equiv A_{\pm a\alpha}^{(b)}=
A_{a\alpha}^{(b)}\pm B_{a\alpha}^{(b)}
\label{eq:A3}
\end{equation}
and
\begin{eqnarray}
{\bf A}^{(3)}&=&A_{+}^{(0)*}C^L A_{+} \, \ \ \ \ \ \ \ \
{\bf E}^{(3)} = A_{-}^{(0)*}C^L A_{+}\nonumber\\
{\bf B}^{(3)}&=& A_{+}^{(0)*}C^L A_{-}\, \ \ \ \ \ \ \ \
{\bf F}^{(3)} = A_{-}^{(0)*}C^L A_{-}\nonumber\\
{\bf C}^{(3)}&=& A_{+}^{(0)*}C^R A_{+}\, \ \ \ \ \ \ \ \
{\bf G}^{(3)} = A_{-}^{(0)*}C^R A_{+}\nonumber\\
{\bf D}^{(3)}&=& A_{+}^{(0)*}C^R A_{-}\, \ \ \ \ \ \ \ \
{\bf H}^{(3)} = A_{-}^{(0)*}C^R A_{-}\,,
\label{eq:A4}
\end{eqnarray}
respectively for those in eqs.(\ref{eq:A11}) and (\ref{eq:A12}).

The UV-divergences of the formulae (\ref{eq:V1},\ref{eq:V2}) are
cancelled by addying
the contribution  from the counterterms in eq.(\ref{eq:RV}) generated by
wave-function renormalization of the external fermions. These
are sketched in Fig.2, where all indices are understood to be summed over.
The (real part of the) self-energy diagram s1 in Fig.2 is given by
$-i\Sigma_{s1}^{(b)}(p)$, with
\begin{eqnarray}
\Sigma_{s1}^{(b)}(p) &=&\left({-ig^2\over 8}\right)
\left\{\right.
[\,|A_{+ a\alpha}^{(b)}|^2\not{p} P_L+|A_{- a\alpha}^{(b)}|^2\not{p} P_R\,]
\,B_1(p^2, M_{\alpha}^0, m_a(\tilde{b}))
\nonumber\\
& - & M_{\alpha}^0\,A_{+ a\alpha}^{*(b)}A_{- a\alpha}^{(b)}
B_0(p^2, m_a(\tilde{b}), M_{\alpha}^0)
\,\left.\right\}\,,
\label{eq:s1}
\end{eqnarray}
where we have used the 2-point function notation $B_{0,1}$
of ref.\cite{b023}.
 From eq.(\ref{eq:s1}), the terms in the decomposition (\ref{eq:Sigma})
immediately read off.
Similarly, $-i\Sigma_{s2}^{(b)}(p)$ from diagram s2 furnishes
\begin{eqnarray}
\Sigma_{s2}^{(b)}(p) &=&\left({-ig^2\over 4}\right)
\left\{\right.
[\,|A_{+ ai}^{(t)}|^2\not{p} P_L+|A_{- ai}^{(t)}|^2\not{p} P_R\,]
\,B_1(p^2, M_{i}, m_a(\tilde{t}))
\nonumber\\
& - & M_{i}\,A_{+ ai}^{*(t)}A_{- ai}^{(t)}
B_0(p^2, m_a(\tilde{t}), M_{i})
\,\left.\right\}\,.
\label{eq:s2}
\end{eqnarray}
As for the diagrams s3 and s4, the contribution from the former follows
 from eq.(\ref{eq:s1}) upon
replacing $A_{\pm a\alpha}^{(b)}\rightarrow A_{\pm a\alpha}^{(t)}$
and $m_a(\tilde{b})\rightarrow m_a(\tilde{t})$, whereas the yield from the
latter drops from
eq.(\ref{eq:s2}) after $A_{+ ai}^{(t)}\rightarrow A_{+ ai}^{(b)}$
and $m_a(\tilde{t})\rightarrow m_a(\tilde{b})$.
Concerning the SUSY contributions to the external $W$-self-energy, they are
shown in Fig.3. We omit the lengthy analytical expressions, which can be
found in ref.\cite{b018}. The same reference also quotes the complete
SUSY contributions
to the self-energies of the Z and of the photon.
We have explicitly checked that when putting everything together,
UV-divergences cancel in eq.(\ref{eq:RV}) and dimensionful logarithms rescale
appropriately. Essential for this are the unitarity of the diagonalizing
matrices $U,V,N$ and $R^{(q)}$ from which all coupling matrices have been
built up.

With all the one-loop SUSY contributions to the form factors identified, the
radiatively corrected amplitude for the process
$t\rightarrow W^{+}\, b$  can be written as follows
($\epsilon^{\mu}$ being the polarization 4-vector of the $W^{+}$):
\begin{eqnarray}
\bar{u}(p_b)\Gamma_{\mu}\,u(p_t)\epsilon^{\mu} &=&
i\,{g\over \sqrt{2}}\left\{\right.[1+F_L+\delta Z_L
+{1\over 2}\hat{\Pi}_t(m_t^2)+
{1\over 2}\hat{\Sigma}_W'(M_W^2)]{\cal M}_0+F_R{\cal M}_1\nonumber\\
&+& H_L{\cal M}_2+H_R{\cal M}_3\left.\right\}\,,
\label{eq:RA}
\end{eqnarray}
where the structure of the reduced matrix elements ${\cal M}_{0,1,2,3}$
should be apparent by comparison
of eqs.(\ref{eq:RV}) and (\ref{eq:RA}). The corrected width now follows after
computing the interference between
the tree-level amplitude and the one-loop amplitude. On the whole we have
\begin{eqnarray}
\Gamma &=& \Gamma_0(\alpha)\,\left\{\right.1+2\,F_L+2\delta Z_L
+\hat{\Pi}_t(m_t^2)+\hat{\Sigma}_W'(M_W^2)\nonumber\\
& + & 2{G_1\over G_0} \,F_R+ 2{G_2\over G_0} \,H_L+ 2{G_3\over G_0} \,H_R
\left.\right\}\equiv\Gamma_0(\alpha)(1+\delta^{SUSY}(\alpha))\,,
\label{eq:width}
\end{eqnarray}
where
\begin{equation}
\Gamma_0(\alpha)=\left({\alpha\over s_{\theta}^2}\right)\,m_t\,|V_{tb}|^2
{G_0\,\lambda^{1/2} (m_t, M_W ,m_b)\over 16 m_t^4}\,,
\label{eq:treealpha}
\end{equation}
with
\begin{equation}
\lambda^{1/2} (x ,y , z)=\sqrt{[x^2-(y+z)^2][x^2-(y-z)^2]}\,,
\end{equation}
is the tree level width, and the polarization sums

\begin{equation}
\begin{array}{ccccl}
G_0 &\equiv &\sum_{pol}|{\cal M}_0|^2
& = & m_t^2+m_b^2-2M_W^2+{(m_t^2-m_b^2)^2\over M_W^2}\,,\nonumber\\
G_1 &\equiv &\sum_{pol}{\cal M}_0{\cal M}_1^{*}
& = & -6m_tm_b\,,\nonumber\\
G_2 &\equiv &\sum_{pol}{\cal M}_0{\cal M}_2^{*}
& = & -{m_t\over M_W}\left[m_t^2+m_b^2-{1\over 2}M_W^2-{(m_t^2-m_b^2)^2
\over 2M_W^2}\right]\,,\nonumber\\
G_3 &\equiv &\sum_{pol}{\cal M}_0{\cal M}_3^{*}
& = & {m_b\over m_t} G_2\,.
\end{array}
\end{equation}

In eq.(\ref{eq:width}), $\delta^{SUSY}(\alpha)$ stands for the
``relative SUSY correction'' in the $\alpha$-scheme,
i.e. the one-loop SUSY correction to the top quark
width with respect to the tree-level width, $\Gamma_0(\alpha)$, in
that scheme.
We emphasize that eq.(\ref{eq:treealpha}) can also
be conveniently
parametrized in terms of $G_F$ (Fermi's constant in $\mu$-decay) by using
\begin{equation}
{\alpha\over s_{\theta}^2}={\sqrt{2}\,G_FM_W^2\over \pi}
(1-\Delta r^{MSSM})\,,
\label{eq:DeltaMW}
\end{equation}
where $s_{\theta}^2$ is given in eq.(\ref{eq:angles}), with the understanding
that $M_W$ and $M_Z$ are the
physical masses of the weak gauge bosons. \, $\Delta r^{MSSM}$
involves all possible radiative corrections, universal (U) and
non-universal (NU) to $\mu$-decay in the MSSM,
in particular the ``genuine'' SUSY ones:
\begin{eqnarray}
 \Delta r^{MSSM} & =& \Delta r^U+\Delta r^{NU}=
-{\hat{\Sigma}_W(0)\over M_W^2}+\Delta r^{NU}\nonumber\\
& = & \Delta r^{SM}+\Delta r^{SUSY}\,,
\label{eq:deltar}
\end{eqnarray}
where $\hat{\Sigma}_W(0)$,
the renormalized self-energy of the $W$ at zero frequency,
is obtained from eq.(\ref{eq:RSEW}).
In the present context, $\Delta r^{SM}$ above includes, apart from
conventional SM physics\,\cite{b110}, also the contribution from the
two-doublet Higgs sector of the MSSM\,\cite{b026}--instead of the single Higgs
doublet of the SM--,
whilst the ``genuine'' SUSY part
is contained in the second term on the RHS of (\ref{eq:deltar}). Clearly,
in the new parametrization $(G_F, M_W, M_Z)$
(call it ``$G_F$-scheme''), the tree-level
width of the top quark, $\Gamma_0(G_F)$,
is related to eq.(\ref{eq:treealpha}) through
\begin{equation}
\Gamma_0(\alpha)=\Gamma_0(G_F)(1-\Delta r)\,,
\label{eq:treeG_F}
\end{equation}
where
\begin{equation}
\Gamma_0(G_F)=\left({G_F M_W^2\over 8\pi\sqrt{2}}\right)\,m_t\,|V_{tb}|^2
{G_0\,\lambda^{1/2} (m_t ,M_W ,m_b)\over m_t^4}\,.
\label{eq:treeGF}
\end{equation}
Hence the ``relative SUSY correction'' with respect to $\Gamma_0(G_F)$
is no longer $\delta^{SUSY}(\alpha)$ but
\begin{equation}
\delta^{SUSY}(G_F)=\delta^{SUSY}(\alpha)-\Delta r^{SUSY}\,.
\label{eq:deltaSUSYG_F}
\end{equation}
The parameters in the $\alpha$-and-$G_F$-schemes are related
by the fundamental relation (\ref{eq:DeltaMW}), in which
$\Delta r^{MSSM}$ plays a crucial role. As for $\Delta r^{SUSY}$
in the MSSM, a full one-loop numerical analysis including all
possible ``genuine'' SUSY (universal, as well as non-universal)
contributions has recently been considered in ref.\cite{b007}
on the basis of an adaptation to the $\alpha$-scheme of
the analytical work of ref.\cite{b018}, which was carried out in a
different (low-energy) renormalization scheme.
We refrain from writing out
the corresponding formulae. These, together with
detailed diagrams contributing
to $\Delta r^U$ and $\Delta r^{NU}$ in eq.(\ref{eq:deltar}),
are displayed in ref.\cite{b018}. We
shall explicitly include these results for the complete
numerical analysis presented in the next section.


\vspace{1.5cm}

\begin{Large}
\begin{quote}
\raggedright
 {\bf 4. Numerical Analysis and Discussion}
\end{quote}
\end{Large}
 \vspace{0.5cm}

The relevant quantities in our analysis are the relative
supersymmetric corrections $\delta^{SUSY}(\alpha)$ and $\delta^{SUSY}(G_F)$
to the top quark width in the $\alpha$-and $G_F$-schemes.
It is well known that in some calculations
it is useful to replace the former scheme with a modified (constrained)
 $\alpha$-scheme based on the parametrization $(\alpha, G_F, M_Z)$\,
\cite{b110}.
Nevertheless, if we would use that framework, then
$M_W$ in  eq.(\ref{eq:treealpha})
would no longer be an input parameter but a (model-dependent) computable
quantity from the constraint  eq.(\ref{eq:DeltaMW}). For this reason we
prefer to stick to the original $\alpha$-scheme, where $M_W$ remains an
input datum. In this respect it is useful to remember that
at LEP 200 the $W$-mass will be measured with a remarkable precision of
$\delta M_W=\pm 28\,(stat.)\pm 24\,(syst.)\, MeV$\,\cite{b027}
\footnote{In ref.\cite{b007} it is shown that a measurement of the $W$-mass
with that precision, or even a factor of two worse, would enable us
to hint at virtual SUSY effects even if the full supersymmetric
spectrum lies in the vicinity of the unaccessible LEP 200 range
(\,$\stackrel{\scriptstyle >}{{ }_{\sim}} 100\,GeV$).}.
Moreover, on top of this it is clear that
for processes dominated by mass scales of order $G_F^{-1/2}$
--as in
our case--it becomes more
appropriate to use the $G_F$-scheme, $(G_F,M_W,M_Z)$, which is a genuine
high energy scheme for electroweak physics. In this parametrization, large
radiative corrections are avoided due to important cancellations between
$\delta(\alpha)$ and $\Delta r$ in eq.(\ref{eq:deltaSUSYG_F}). This is
a reflection of the well-known
fact that $G_F$ (as extracted from $\mu$-decay) does not run from
low-energy up to the electroweak scale, since large logarithms
associated to the renormalization group (RG) do not show up.

Although we shall compare in some respects the radiative corrections
in the $\alpha$-and $G_F$-schemes,
we present the bulk of our numerical analysis in the $G_F$-scheme.
The actual corrections
can be straightforwardly computed upon making
use of the explicit formulae from sect.3
and the analysis of $\Delta r^{SUSY}$ from ref.\cite{b007}.
In practice, however, the numerical evaluation of these formulas is
technically non-trivial since it
requires exact treatment of the various 2 and
3-point functions for nonvanishing masses and external momenta
\footnote{Leading order calculations performed
in the limit of $m_t$ larger than any other mass scale in the decay process
have proven to fail in
previous analyses within the context of the SM\,\cite{b028,b013}.}.
In particular, the exact evaluation of each scalar 3-point function $C_0$
involves a cumbersome representation
in terms of twelve (complex) Spence functions.
We refer the reader to the standard techniques in the literature\,
\cite{b023,b024,b025}
and go directly to present and discuss the final numerical results.
They are displayed in Figs.4-8. Apart from the basic input parameters
$\alpha$ and $G_F$, we have fixed\,\cite{b029}
\begin{equation}
M_Z=91.187\, GeV\,\,\,,\;\;\;\;\;\;\;\;\;\;\; m_b=4.7\,GeV\,,
\;\;\;\;\;\;\;\;\;\;\;V_{tb}=0.999\,.
\end{equation}
As for the sparticle masses, they have been required
to respect the current phenomenological bounds.
 On  general grounds, the
model-independent bounds from $Sp\bar{p}S$ and LEP are the following
\,\cite{b031}
\begin{equation}
m_{\tilde{l}^{\pm}}\geq 45\,GeV\,,\ \ \ m_{\tilde{\nu}}\geq 42\,GeV\,,
\ \ \ M(\Psi^{\pm})\geq 47\,GeV\,, \ \ \ M(\Psi^{0})\geq 20\,GeV\,.
\label{eq:boundLC}
\end{equation}
Concerning squarks, the absolute LEP limits are, in principle,
similar to those for sleptons\,\cite{b032}.
On the other hand the $Sp\bar{p}S$ searches for
squarks and gluinos ($\tilde{g}$) amount to a more stringent
bound of $m_{\tilde{q}}\geq 74\,GeV$
for $m_{\tilde{g}}$ around $ 80\,GeV$\,
\cite{b031}.
 All the same, this limit
becomes poorer as soon as gluinos become heavier. A similar situation occurs
for the Tevatron limits, which improve the squark mass lower bound up to
$m_{\tilde{q}}\geq 130\,GeV$ for $m_{\tilde{q}}\leq m_{\tilde{g}}\leq
400\,GeV$, but if one permits the gluino masses to go beyond $400\,
GeV$ the squark mass limit disappears\,\cite{b031}.
  We shall not consider this extreme
possibility. Nevertheless since we want to maximize the possible
effects from squarks, we will commence on assuming a mixed mass scenario in
which the following limit on squark masses of the first two generations apply:
\begin{equation}
m_{\tilde{q}}\geq 130\, GeV\;\;\;\;\; (\tilde{q}=\tilde{u},\tilde{d}\,,
\tilde{c},\tilde{s})\,,
\label{eq:boundUD}
\end{equation}
while we shall explore sbottom and stop squarks with masses starting lower
limits
\begin{equation}
m_{\tilde{q}}\geq 75\, GeV\;\;\;\;\; (\tilde{q}=\tilde{t},\tilde{b})\,.
\label{eq:boundTB}
\end{equation}
We will eventually increase this limit up to the
typical bound (\ref{eq:boundUD}) ascribed to the other families.
The reason to single out the third generation of squarks is because
the effects of L-R mixing in the mass matrices
(specially for the stop, but also for the sbottom)
could substantially lower one of the mass eigenvalues (see later on).
Furthermore, the
squarks of the third generation are those
directly involved in the top decay diagrams in Figs.1-2, while the
first two generations of squarks only enter through
the universal (so-called {\sl oblique}\,\cite{b033}) type
corrections shown in Fig.3 --and corresponding ones for the
photon and the $Z$-boson.  The numerical analysis shows that these universal
contributions to $\Gamma_t$,
which are generated by the term $\hat{\Sigma}_W'(M_W^2)$
in eq.(\ref{eq:width}), are rather insensitive to whether we consider
the bound (\ref{eq:boundUD}) or the bound (\ref{eq:boundTB}) for
all sfermions.
It should also be pointed out, in connection to what has been stated above,
 that the quantity
$\delta Z_2^W$ (see eq.(\ref{eq:deltaZ})), which is sensitive
to the RG-running of $\alpha$,
as well as to the mass
splitting among the $T^3=\pm 1/2$ components
in any given $SU(2)_L$ doublet,
turns out to cancel from $\delta^{SUSY}(G_F)$, due to the
difference between $\delta (\alpha)$ and $\Delta r$ in
eq.(\ref{eq:deltaSUSYG_F})
\footnote{Contrary to all light fermions in the SM, the bounds on
sfermion masses given above show that virtual effects
 from squarks and sleptons must
decouple from the photon, as it is also the case for the top quark.
Thus no leading RG-corrections from SUSY particles are
to be expected in the MSSM, not even in the $\alpha$-scheme.}.
Therefore, in the $G_F$-scheme
one expects neither leading RG-type corrections nor
any significant enhancement from
custodial symmetry-breaking contributions induced by large deviations of
the $\rho$-parameter from unity
\footnote{As a matter of fact, custodial symmetry in the MSSM cannot be
broken by non-decoupling universal effects,
whether statical ($\rho$-parameter) or dynamical (wave-function
renormalization of the gauge bosons).
In the limit of $M_{SUSY}\rightarrow\infty$, these
effects must vanish\,\cite{b034},
irrespective of the parametrization.}. The only hope lies in the non-oblique
radiative corrections caused by enhanced Yukawa couplings of the form
(\ref{eq:Yukawas}), and this is precisely what we are after.

 From the point of view of model-building, we have generated the
pattern of sfermion masses preserving the bounds
(\ref{eq:boundLC}), (\ref{eq:boundUD})
and (\ref{eq:boundTB}) by using models with radiatively induced
breaking of the $SU(2)_L\times U(1)_Y$ symmetry such as Supergravity
inspired models
\,\cite{b005}.
 For sleptons and the  first two generations of squarks we have, using the
notation of sect.2,
\begin{equation}
m^2_{{\tilde{f}}_{L,R}}=m^2_f+M^2_{{\tilde{f}}_{L,R}}\pm
\cos{2\beta}\,(T^3_{L,R}-Q_{\tilde{f}}\,\,s_{\theta}^2)\,M_Z^2\,,
\label{eq:sferm}
\end{equation}
where $T^3_{L,R}$ and $Q_{\tilde{f}}$ stand, respectively,
for the third component of weak isospin and electric charge
corresponding to each member of the multiplet and for each ``chiral'' species
$\tilde{f}_{L,R}$ of sfermion.
 Finally, the parameters
$M_{{\tilde{f}}_{L,R}}$ are soft SUSY-breaking mass terms\,\cite{b005}.
The mass splitting between the $T^3=+1/2$ and the $T^3=-1/2$ components in
each $SU(2)_L$ doublet is independent of $M_{{\tilde{f}}_L}$
\begin{equation}
m^2_{\tilde{f}_L(T^3=+1/2)}
-m^2_{\tilde{f}_L(T^3=-1/2)}=M_W^2\,\cos{2\beta}\,,
\label{eq:splitting1}
\end{equation}
where we have neglected the fermion masses squared
 of the first two generations against
the term on the RHS of eq.(\ref{eq:splitting1}). The situation
for the stop-sbottom doublet, however, requires a particular treatment,
due to the possibility of large LR-mixing. We assume it to be the case for
the stop squark and proceed to probe this effect in terms
 of the mass parameter $M_{LR}$ in
the stop mass matrix, which can be written as follows:
\begin{equation}
{\cal M}_{\tilde{t}}^2 =\left(\begin{array}{cc}
M_{\tilde{b}_L}^2+m_t^2+\cos{2\beta}({1\over 2}-
{2\over 3}\,s_{\theta}^2)\,M_Z^2
 &  m_t\, M_{LR}\\
m_t\, M_{LR} &
M_{\tilde{t}_R}^2+m_t^2+{2\over 3}\,\cos{2\beta}\,s_{\theta}^2\,M_Z^2
\end{array} \right)\,.
\label{eq:stopmatrix}
\end{equation}
Here we have used the fact that $SU(2)_L$-gauge invariance requires
$M_{\tilde{t}_L}=M_{\tilde{b}_L}$ and thus the first entry of the matrix can
be writen in terms of the L-sbottom mass parameter $m_{\tilde{b}_L}$.
To illustrate the
effect of the mixing it will suffice to choose the soft SUSY-breaking mass
$M_{\tilde{t}_R}$ in such a way that the two
diagonal entries of  ${\cal M}_{\tilde{t}}^2$ are equal--
the mixing angle is thus fixed at $\pi/4$. In fact,
we have checked that our results are not significantly
 sensitive to large
variations of $M_{\tilde{t}_R}$.
 For the other sfermions
we assume that the R-type and L-type species are degenerate in mass.
In this way the only
two free parameters are  $m_{\tilde{b}_L}$ and $M_{LR}$. For the
mixing parameter, however, we have the proviso
\begin{equation}
M_{LR}\leq 3\,m_{\tilde{b}_L}\,,
\label{eq:MLR}
\end{equation}
which roughly corresponds to a well-known
necessary, though not sufficient, condition to avoid false vacua, i.e. to
guarantee that the $SU(3)_c\times U(1)_{em}$ minimum is the deepest one
\,\cite{b134}.
 Finally, we have also imposed the condition that for our
choices of the parameters the induced
deviations of the $\rho$-parameter from 1 should satisfy the bound
\footnote{For relatively recent detailed discussions on the $\rho$-parameter
in SUSY, see ref. \cite{b036}. The bound (\ref{eq:rho}) is based on pure
SM physics and it could be somewhat more relaxed in the framework
of the MSSM, as it is exemplified in ref.\cite{b037}.} \,\cite{b237}
\begin{equation}
|\delta\rho|\leq 0.005\,.
\label{eq:rho}
\end{equation}

We now come to the discussion of our numerical analysis.
In Figs.4-7 we fix $M_{LR}=0$ and postpone the case of nonvanishing
stop mixing until Fig.8.
As in ref.\cite{b014}, we restrict ourselves
to the following interval of $\tan{\beta}$:
\begin{equation}
1\leq\tan{\beta}\leq 70\,.
\label{eq:tanbeta}
\end{equation}
 In Fig.4 we display
contour lines of $\delta^{SUSY}(G_F)$ in the higgsino-gaugino parameter
space $(\mu, M)$.
 Since we want to compare our SUSY maximum results with those of the
2HDM from refs.\cite{b014,b022}, we have fixed
the value of $\tan{\beta}$ at the upper limit
of the interval (\ref{eq:tanbeta}),
which roughly corresponds to the perturbative limit\footnote{The
lower bound in eq.(\ref{eq:tanbeta}) arises from consistency in GUT models.
In this respect, there are phenomenological indications\,\cite{b137}
and also recent $SO(10)$ unification models that tend to
favor large values of $\tan{\beta}$\,\cite{b038}. Proton decay, however,
gives $\tan{\beta}\leq 85$\,\cite{b039}.}. For very large
($\geq m_t/m_b$) values of $\tan{\beta}$,
the two Yukawa couplings (\ref{eq:Yukawas}) are
in the relation $1<\lambda_t< \lambda_b$,
and therefore they both give sizeable contributions
which translate into relatively large
(negative) corrections $\delta^{SUSY}(G_F)=-(5-10)\%$
on the top quark width
\footnote{For $\tan{\beta}<1$, $\lambda_b$ is very small, but $\lambda_t$
can become rather large and in this region of parameter space one may
recover $\delta^{SUSY}(G_F)=-(5-10)\%$ too. Nevertheless, values of
$\tan{\beta}$ less than one are disfavoured on several
accounts both theoretically and phenomenologically\,\cite{b139}.}.
 This is in contradistinction to the maximally expected
quantum corrections from the Higgs sector of
the MSSM, in which even for $\tan{\beta}=70$ the
Higgs correction is of only $1\%$\,\cite{b014,b022}.
Notice that for large $M$ and $\mu$ the chargino-neutralino
contributions in Fig.4 die away, as expected from the decoupling theorem
\,\cite{b034}.
On Tables I and II we may appreciate more closely
the numerical differences between the corrections
in the $\alpha$-and $G_F$-schemes for a few choices of the SUSY parameters
$M,\mu$ and $m_{\tilde{f}}$. The corresponding induced value of
$\delta\rho ^{SUSY}$ is also provided and it is seen to preserve the bound
(\ref{eq:rho}).
On the other hand, in Fig.5 (i) we may assess the dependence
of  $\delta^{SUSY}(G_F)$ on $\tan{\beta}$ in the full range (\ref{eq:tanbeta})
for typical choices of the other parameters. The corresponding (larger)
corrections in the $\alpha$-scheme are seen in Fig.5 (ii).

We remark that in Fig.4 we have also explicitly displayed
the  singular contour
lines corresponding to the threshold singularities that are expected from the
derivatives of the renormalized self-energy functions of the top quark and of
the $W$-gauge boson
(Cf. eqs. (\ref{eq:pito}) and (\ref{eq:DRSEW})) in conventional
perturbation theory \,\cite{b040}.
 For a  fixed value of $\tan{\beta}$ and of the sfermion masses,
$m_{\tilde{f}}$, the singularities from wave-function renormalization
appear for every numerical pair $(\mu, M)$ for which the
corresponding eigenvalues of the chargino and neutralino mass
matrices satisfy one of the following relations
\begin{equation}
m_t=M_i+m_{\tilde{b}}=M_{\alpha}^0+m_{\tilde{t}}\;,\;\;\;\;\;\;\;
M_W=M_i+M_{\alpha}^0=m_{\tilde{t}}+m_{\tilde{b}},.
\label{eq:singular}
\end{equation}
Those (fake) singularities are extremely well concentrated around the set
of points satisfying eqs. (\ref{eq:singular}). By explicitly plotting the
threshold isolines, it becomes patent in Fig.4 that
the typical contours $\delta^{SUSY}(G_F)=-(4, 6) \%$
do not cross the troublesome
(pseudo-) singularities. Similarly, the prominent
spikes standing up in Figs.5-8
correspond to the projection of the threshold effects onto the different
SUSY 2-parameter spaces selected in our analysis.

 Several techniques have been
deviced to tackle this problem-- which is not new
and is also encountered in other contexts and
in particular in pure SM physics\,\cite{b041}:
One may e.g. resort to
appropriate
Dyson-resummation of the propagator of the unstable particle (in our case the
top quark and the $W$-boson) so that the
derivative of its self-energy appears in the denominator, or
alternatively, one may define the
mass and width of the unstable particle through the complex pole position
of its propagator, thus avoiding explicit wave function renormalization
\,\cite{b040}.
In practice, however, since such spurious effects are strongly
localized and there is no unambiguous recipe to interpolate the perturbative
behaviour, we can get rid of them either by removing the immediate
neighbourhood around these points from our numerical analysis or simply
by explicitly including these narrow domains but not trusting
the results from the inside points, where the perturbative expansion of the
$S$-matrix elements of the theory breaks down. The latter procedure
is in actuality our own approach.

The mass of the top quark is, together with a large value of $\tan{\beta}$,
another enhancement factor for our radiative corrections.
In Fig.6 (i) we show the behaviour of $\delta^{SUSY}(G_F)$
in terms of the top
quark mass in the range $100\leq m_t \leq 200\, GeV$. After crossing the
transient threshold effects, corrections of order $-10\%$ can be achieved.
In Fig.6 (ii) we exhibite for comparison the
corresponding corrections in the $\alpha$-scheme.
In order to keep an eye on
the bound (\ref{eq:rho}), we plot $\delta\rho^{SUSY}$ as a
function of the  top quark mass in Fig.6 (iii). We see that for the cases
under consideration the bound is saturated at
$m_t\simeq 170\,GeV$.
Finally, in Figs.7-8 we investigate
the dependence of the correction on the mass of the sbottom and stop squarks
in the $G_F$-scheme.
Remember that
in our framework these masses are fixed in terms of the two parameters
$m_{\tilde{b}_L}$ and $M_{LR}$. For $M_{LR}=0$, Fig.7 shows the behaviour
of the correction against $m_{\tilde{b}}$.
For large enough values of
this parameter (and therefore of the sbottom and stop masses), the correction
decreases (decoupling). However, at the boundary values of the
central interval
$75\,GeV< m_{\tilde{b}_L}< 130\, GeV$ ( where two of the
singular--removable-- spikes show up
on the middle) the correction remains fairly the same . Thus we learn that
it does not make much difference to consider stop-sbottom masses near
the original lower bound (\ref{eq:boundTB}) or near the more conservative
lower bound
(\ref{eq:boundUD}) that we assumed for the other squark families.
This conclusion does not change when we switch on the mixing parameter
$M_{LR}$. It is true that for large values of $M_{LR}$ there emerges a light
eigenvalue of the stop mass matrix (\ref{eq:stopmatrix}), as seen in
Fig.8 (i). All in all the bare fact is that
due to a partial cancellation among vertices and external
self-energies (reminiscent of a Ward identity), in combination to
the aforementioned
cancellation of $\delta Z_2^W$ from $\delta^{SUSY}(G_F)$,
the possibility of having a large splitting between one light stop
and a heavy sbottom does not render any substantial correction
to $\Gamma_t$. In Fig.8 (ii) we confirm that outside the singular spikes the
correction is not very sensitive (typically $\leq 10\%$) to $M_{LR}$.

In conclusion, there could be relatively large (few to 10 percent)
non-oblique electroweak corrections
to the top quark width from the ``genuine'' SUSY part of the MSSM, due to
enhanced Yukawa couplings in the gaugino-higgsino sector. This is in
contrast to the one-doublet Higgs sector of the SM, and also to
the two-doublet Higgs sector of the MSSM, where in comparable conditions the
corrections are one order of magnitude smaller. It is also remarkable that
the supersymmetric electroweak corrections are of the same
(negative) sign and could
be of the same order of magnitude than the conventional QCD corrections
\,\cite{b111,b211,b013}. However, whereas
the latters are almost insensitive to the
top quark mass in the wide range $130\,GeV\le m_t\le 300\,GeV$,
the formers do significantly vary with $m_t$ in the narrow
relevant range $150\,GeV\le m_t\le 200\,GeV$.
On the whole the QCD + SUSY corrections could reduce
the top quark width up to about $10-20\%$. Consequently, a measurable
reduction beyond $\simeq 8\%$ (QCD)
could be attributted to a ``genuine'' SUSY effect.
The fact that the gaugino-higgsino sector of the MSSM could afford a
non-negligible quantum correction to the top quark decay width,
in contradistinction to the inappreciable
yield from the scalar Higgs sector of the MSSM, can be traced to the
highly constrained structure of the Higgs potential as dictated by SUSY.

Two final remarks:\, i) In the MSSM, the decay $t\rightarrow W^{+}\,b$ is
not the only possible decay; there could be additional electroweak
decay modes, such as $t\rightarrow H^{+}\,b$ and
$t\rightarrow \tilde{t}\,\Psi_{\alpha}^0$, and they have been studied in
detail \,\cite{b042}. These SUSY modes notwithstanding,
the canonical decay channel would always give a
large branching ratio. And in the event that the new decay channels would be
closed, due to phase space (i.e. for heavy enough charged
Higgs and stop),
our SUSY corrections to the
the canonical decay could still remain sizeable (Cf. Fig.7);\,
ii)  In this work, we have not addressed the computation of the strong
SUSY corrections to the top quark width, since one usually assumes that
gluinos are very heavy and therefore give negligible contributions.
Nonetheless this conclusion could change dramatically
if one takes seriously the possibility that
gluinos could be light (few $GeV$) \,\cite{b043}
 or relatively light ($\simeq 80-90\,GeV$) \,\cite{b044,b045}. In those
cases a new SUSY channel, $t\rightarrow \tilde{t}\,\tilde{g}$, could be open
and compete with the canonical mode. Alternatively, it could be closed, but
the strong SUSY radiative corrections to the canonical decay be rather
significant.
It would certainly be interesting to investigate the impact of a light
(or relatively light ) gluino
scenario on the top quark width, but this goes beyond the
scope of the present work\,\cite{b046}.

\vspace{1.25cm}

{\bf Acknowledgements}: One of us (JS) is thankful to the Spanish
Ministerio de
Educaci\'on y Ciencia for finantial support through the program of Acciones
Integradas and is also grateful to the hospitality of the
Max Planck Institut in Munich where
part of this work was carried out.
He has also benefited from conversations with
M. Mart\'{\i}nez, F. Paige, P. Sphicas and P. Zerwas, and is indebted to
R. Miquel as well as to M. Carena and C. Wagner
for providing a useful reference. The work of DG, RJ and JS has
been partially supported by CICYT under project No. AEN93-0474.



\newpage
\begin{center}
\begin{Large}
{\bf Figure Captions}
\end{Large}
\end{center}
\begin{itemize}
\item{\bf Fig.1} Feynman diagrams, up to one-loop order, for the SUSY
vertex corrections to the top quark decay process $t\rightarrow W^{+}\,b$.
Each one-loop diagram is summed over all possible values of the
mass-eigenstate charginos ($\Psi^{\pm}_i\,; i=1,2$), neutralinos
($\Psi^{0}_{\alpha}\,; \alpha=1,2,...,4$), stop and sbottom squarks
($\tilde{b}_a, \tilde{t}_b\,; a,b=1,2$).

\item{\bf Fig.2} One-loop SUSY contributions to the external
fermion self-energies in the decay process $t\rightarrow W^{+}\,b$. The
notation is as in Fig.1.

\item{\bf Fig.3}\, SUSY vacuum polarization effects on the gauge boson
$W^{+}$. Chargino-neutralino notation as in Fig.1. All six sfermion
families contribute, whether sleptons ($\tilde{f}=\tilde{e},...,\tilde{\tau};
\tilde{f}'=\tilde{\nu}_e,...,\tilde{\nu}_{\tau}$) or squarks
($\tilde{f}=\tilde{u},...,\tilde{t}; \tilde{f}'=\tilde{d},...,\tilde{b}$).

\item{\bf Fig.4} Typical contour (dotted) lines ${\delta}^{SUSY}
(G_F)=-(4,6) \%$ in the higgsino-gaugino
$(\mu,M)$-parameter space for $\tan{\beta}=70$ and
$m_t=160\, GeV$. We have fixed $m_{\tilde{\nu}_f}=50\,GeV$ and the lower
bounds in eqs.(\ref{eq:boundUD}) and
(\ref{eq:boundTB}) in combination with the model relation
eq.(\ref{eq:sferm}). The dashed $t$-and $W$-lines correspond to the
threshold pseudo-singularities associated to the wave-function
renormalization of the $t$-quark and $W$-boson, respectively.
The blank regions delimited by the full lines
are phenomenologically excluded by the
constraints  $ M(\Psi^{\pm})\geq 47\,, M(\Psi^{0})\geq 20\,GeV $. Points
$a$, $b$ and $c$ are used to fix $(\mu, M)$ in Figs.5-8.

\item{\bf Fig.5} Dependence on $\tan{\beta}$ of
(i) ${\delta}^{SUSY}(G_F)$ and (ii) ${\delta}^{SUSY}(\alpha)$ for three
widely different choices of the higgsino-gaugino parameters $(\mu,M)$:
 $(-100, 100)$ (curve a),
$(-180,120)$ (curve b) and $(-60, 200)$ (curve c).
Remaning parameters as in Fig.4.

\item{\bf Fig.6} Dependence on $m_t$ of (i) ${\delta}^{SUSY}(G_F)$,
(ii) ${\delta}^{SUSY}(\alpha)$ and (iii)
$\delta\rho^{SUSY}$,  for the same choices of sfermion
and higgsino-gaugino parameters, $(\mu,M)$, as in Fig.5. In Fig.6 (iii),
the cases $(-180, 120)$ and $(-60, 200)$ are almost indistinguishable and
they are both represented by the line $b\simeq c$.

\item{\bf Fig.7} The correction ${\delta}^{SUSY}(G_F)$ as a function of the
sbottom mass. All other masses are fixed as in Fig.4. The three curves
correspond to values of $(\mu, M)$ as in Fig.5.

\item{\bf Fig.8} (i) The evolution of the light and heavy stop masses as
a function of the mixing parameter $M_{LR}$; (ii) Variation of
${\delta}^{SUSY}(G_F)$ in terms of $M_{LR}$ for the three choices of
$(\mu,M)$ as in Fig.5.
\end{itemize}


\newpage
\begin{center}
\begin{Large}
{\bf Table Captions}
\end{Large}
\end{center}
\begin{itemize}
\item{\bf Table I.}\, Numerical comparison of the radiative corrections
\,${\delta}^{SUSY}(G_F)$ and
 ${\delta}^{SUSY}(\alpha)$ for a few choices
of chargino-neutralino masses around points a, b and c defined
in Fig.4. We
 have fixed the remaining parameters also as in Fig.4; in particular,
 $m_{\tilde{b}}=75\,GeV$. On the first column of the table we include the
induced value of $\delta\rho^{SUSY}$. All numbers are given in percent.

\item{\bf Table II.}\, As in Table I, but with $m_{\tilde{b}}=130\,GeV$.

\end{itemize}


\newpage
\thispagestyle{empty}
\begin{table}[p]
\centering

\begin{tabular}{|c|ccc|}
  \multicolumn{4}{c}{\hfill} \\
  \multicolumn{4}{c}{\bf $m_{\tilde{b}}\ =\ 75\ GeV$} \\
  \multicolumn{4}{c}{\hfill} \\
  \hline
  & \multicolumn{3}{|c|}{\hfill} \\
  $(\mu,M)$ & $\delta {\rho}^{SUSY}$ & ${\delta}^{SUSY} (\alpha)$
        & ${\delta}^{SUSY} (G_F)$ \\[.10cm] \hline
  & \multicolumn{3}{|c|}{\hfill} \\
  & & & \\

  ($\;$-180,$\;$110)  &   0.437    &   -7.27   &   -5.10 \\
  ($\;$-180,$\;$120)  &   0.436    &   -5.75   &   -3.59 \\
  ($\;$-180,$\;$130)  &   0.434    &	-5.16   &   -3.01 \\

  ($\;$-100,\ \ 90)   &   0.474    &   -9.69   &   -7.21 \\
  ($\;$-100,$\;$100)  &   0.469    &   -9.33   &   -6.91 \\
  ($\;$-100,$\;$110)  &   0.464    &   -8.66   &   -6.28 \\

  (\ \ -60,$\;$190)  &   0.443    &  -11.27   &   -9.05 \\
  (\ \ -60,$\;$200)  &   0.439    &  -11.51   &   -9.29 \\
  (\ \ -60,$\;$210)  &   0.436    &  -11.57   &   -9.36 \\
\hspace{3.5cm}  & \hspace{2.5cm} & \hspace{2.5cm} & \hspace{2.5cm} \\
\hline
\end{tabular}

\vspace{0.5cm}
{\bf Table I.}
\end{table}

\begin{table}[p]
\centering

\begin{tabular}{|c|ccc|}
  \multicolumn{4}{c}{\hfill} \\
  \multicolumn{4}{c}{\bf $m_{\tilde{b}}\ =\ 130\ GeV$} \\
  \multicolumn{4}{c}{\hfill} \\
  \hline
  & \multicolumn{3}{|c|}{\hfill} \\
  $(\mu,M)$ & ${\delta} {\rho}^{SUSY}$ & ${\delta}^{SUSY} (\alpha)$
        & ${\delta}^{SUSY} (G_F)$ \\[.10cm] \hline
  & \multicolumn{3}{|c|}{\hfill} \\
  & & & \\

  ($\;$-180,$\;$110)  &   0.310   &   -3.24   &   -1.65 \\
  ($\;$-180,$\;$120)  &   0.309   &   -3.13   &   -1.56 \\
  ($\;$-180,$\;$130)  &   0.307   &   -3.05   &   -1.48 \\

  ($\;$-100,\ \ 90)   &   0.347   &   -7.53   &   -5.63  \\
  ($\;$-100,$\;$100)  &   0.342   &   -6.70   &   -4.86 \\
  ($\;$-100,$\;$110)  &   0.337   &   -6.21   &   -4.41 \\

  (\ \ -60,$\;$190)  &   0.316   &  -10.47   &   -8.83 \\
  (\ \ -60,$\;$200)  &   0.312   &   -9.94   &   -8.31 \\
  (\ \ -60,$\;$210)  &   0.309   &   -9.55   &   -7.92 \\
\hspace{3.5cm}  & \hspace{2.5cm} & \hspace{2.5cm} & \hspace{2.5cm} \\
\hline
\end{tabular}

\vspace{0.5cm}
{\bf Table II.}
\end{table}


\end{document}